\documentclass{aipproc}
\usepackage{aipxfm}
\layoutstyle{6x9}
\begin{document}

\title
[Neutrino Self-energy]
{\bf Neutrino self-energy in a magnetized 
charge-symmetric medium}


\author{Alberto  Bravo Garc\'{i}a}{
  address={Nacional Autonoma de Mexico, Circuito Exterior, C.U.,
  A. Postal 70-543, C.P. 04510 Mexico DF, Mexico},
}

\iftrue
\author{Kaushik Bhattacharya}{
address={Nacional Autonoma de Mexico, Circuito Exterior, C.U.,
A. Postal 70-543, C.P. 04510 Mexico DF, Mexico},
}

\author{Sarira Sahu}{
address={Nacional Autonoma de Mexico, Circuito Exterior, C.U.,
A. Postal 70-543, C.P. 04510 Mexico DF, Mexico},
}
\fi

\begin{abstract}
In this talk we present the calculation of the neutrino self-energy in
presence of a magnetized medium. The magnetized medium consists of
electrons, positrons, neutrinos and a uniform classical magnetic
field. The background magnetic field is assumed to be weak compared to
the $W$-Boson mass as a consequence of which only linear order
corrections in the field are included in the $W$ boson propagator. The
electron propagator consists all order corrections in the background
field. Our calculation is specifically suited for situations where the
background plasma may be CP symmetric. 
\end{abstract}

\maketitle
\section{Introduction}
\label{intro}
In the present talk we assume the magnetic field to be less than the
critical field corresponding the $W^{\pm}$ bosons and consequently we
take only linear order, in the magnetic field strength, corrections to
its propagator. The gauge bosons are assumed to be not in thermal
equilibrium and so their thermal modifications are not used. It is
seen that the neutrino self-energy to linear order in Fermi coupling,
$G_F$ vanishes when the number of particles equals the number of
anti-particles in the plasma. In astrophysical cases this does not
happen in general as at these temperatures rarely particle numbers
equals anti-particle numbers. But in the early universe and probably
in the GRB fireball the temperature were supposed to be very high and
so at these places we can expect that the number of particles equaled
that of the antiparticles, to a great extent. In these circumstances
we show that only order $G_F^2$ contributions remain in the expression
of the neutrino self-energy.  We have worked in the unitary gauge and
have not discussed about the gauge independence of the result, as it
is noted that in such calculations the self-energy generally is
dependent on the gauge choice but the dispersion relation is
independent of the gauge \cite{Erdas:1990gy, Erdas:2000iq}.
\section{General expression for the neutrino self-energy in a magnetized 
charge symmetric medium }
\label{renexp}
The effect of the medium is represented by the 4-velocity
of its centre-of-mass $u^\mu$ which looks like:
\begin{eqnarray}
u^\mu = (1, {\bf 0})\,,
\label{u}
\end{eqnarray}
in the rest frame of the medium. The $u^\mu$ is normalized in such a way that,
\begin{eqnarray}
u^\mu u_\mu =1\,.
\label{unorm}
\end{eqnarray} 
Likewise the effect of the magnetic field enters through the 4-vector
$b^\mu$ which is defined in such a way that the frame in which the
medium is at rest,
\begin{eqnarray}
b^\mu = (0, \hat{\bf b})\,,
\label{b}
\end{eqnarray}
where we denote the magnetic field vector by ${\mathcal B} \hat{\bf
b}$. The 4-vector $b^\mu$ is defined in such a way that,
\begin{eqnarray} 
b^\mu b_\mu = -1\,.
\label{bnorm}
\end{eqnarray}
In this talk we take the background classical magnetic field vector to
be along the $z$-axis and consequently $b^\mu=(0,0,0,1)$. The most
general form of the neutrino dispersion relation in the magnetized
plasm is of the form:
\begin{eqnarray}
\Sigma (k) &=& R \Big( a_\parallel k^\mu_\parallel + a_\perp k^\mu_\perp 
+ b u^\mu + c b^\mu \Big) \gamma_\mu L \,,
\label{sigmanb}
\end{eqnarray}
where $k^\mu_\parallel=(k^0, k^3)$ and $k^\mu_\perp=(k^1, k^2)$.  With
the above form of the neutrino self-energy the dispersion relation
comes out as:
\begin{eqnarray}
(1-a_\parallel)E_{\nu_\ell} = \pm \left[\left((1-a_\parallel)k_3 + c\right)^2
+ (1-a_\perp)^2 k_\perp^2\right]^{1/2} + b\,.
\label{mndisp}
\end{eqnarray}
The coefficients $a$, $b$ and $c$ are functions of $k^2_\parallel$,
$k^2_\perp$, $k \cdot u$ and $k \cdot b$. The results of the
calculation are written in the rest frame of the plasma where the
magnetic field points in the $z$-direction of the coordinate system.
\section{The form of the neutrino self-energy}
\label{dnsn}
\begin{figure}[h!]
{\includegraphics{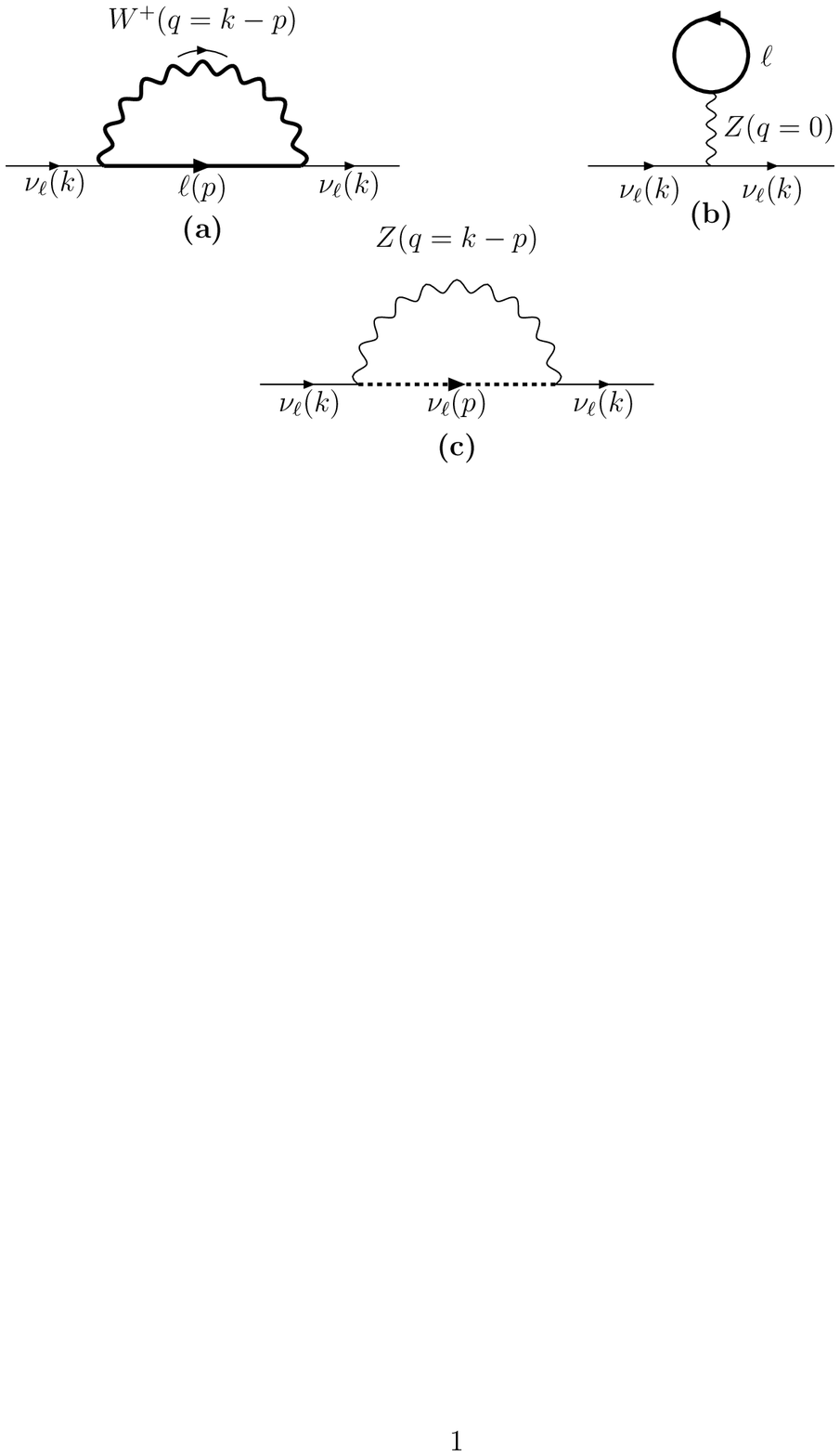}}
\caption{\sf One-loop diagrams for neutrino self-energy in a
magnetized medium. Diagrams (a) and (c) are the $W$ and $Z$ exchange diagrams 
and diagram (b) is the tadpole diagram. The heavy internal lines in diagrams 
(a) and (b) represent the $W$ and the charged lepton propagators in a 
magnetized medium.  
The heavy dashed internal neutrino line in diagram (c) corresponds to the 
neutrino propagator in a thermal medium.
\label{f:selfen1}}
\end{figure}

In this talk we will work in the unitary gauge and consequently the
three diagrams corresponding to the neutrino self-energy are as given
in Fig.~\ref{f:selfen1}.
The one-loop neutrino self-energy in a magnetized medium is comprised
of three pieces, one coming from the $W$-exchange diagram which we
will call $\Sigma^W (k)$, one from the tadpole diagram which we will
designate by $\Sigma^T (k)$ and one from the $Z$-exchange diagram
which will be denoted by $\Sigma^Z (k)$. The total self-energy of the
neutrino in a magnetized medium then becomes:
\begin{eqnarray}
\Sigma(k) = \Sigma^W(k) + \Sigma^T (k) + \Sigma^Z (k)\,.
\label{tsen}
\end{eqnarray}
Here 
\begin{eqnarray}
-i\Sigma^W(k)=\int\frac{d^4 p}{(2\pi)^4}\left(\frac{-ig}{\sqrt{2}}\right)
\gamma_\mu\, L \,iS_{\ell}(p)\left(\frac{-ig}{\sqrt{2}}\right)\gamma_\nu 
 \,L\,i W^{\mu \nu}(q)\,,
\label{wexch}
\end{eqnarray}
\begin{eqnarray}
-i\Sigma^T(k)= -\left(\frac{g}{2\cos \theta_W}\right)^2 R\,
\gamma_\mu\,iZ^{\mu \nu}(0)\int\frac{d^4 p}{(2\pi)^4} {\rm Tr}
\left[\gamma_\nu \,(c_V + c_A \gamma_5)\,iS_{\ell}(p)\right]\,,   
\label{tad}
\end{eqnarray}
and
\begin{eqnarray}
-i\Sigma^Z(k)=\int\frac{d^4 p}{(2\pi)^4}\left(\frac{-ig}
{\sqrt{2}\cos\theta_W}\right)
\gamma_\mu\, L \,iS_{\nu_\ell}(p)\left(\frac{-ig}{\sqrt{2}\cos\theta_W}\right)
\gamma_\nu\,L\,i Z^{\mu \nu}(q)\,.
\label{Zexch}
\end{eqnarray}
In the above expressions $g$ is the weak coupling constant and
$\theta_W$ is the Weinberg angle.  The quantities $c_V$ and $c_A$ are
the vector and axial-vector coupling constants
which come in the neutral-current interaction of
electrons, protons ($p$), neutrons ($n$) and neutrinos with the $Z$
boson.  Their forms are as follows,
\begin{eqnarray}
c_V=\left \{\begin{array}
{r@{\quad\quad}l}
-\frac{1}{2}+2\sin^2\theta_W & e\\
\frac{1}{2} & {\nu_e}\\ \frac{1}{2}-2\sin^2\theta_W & {{p}}\\
-\frac{1}{2} & {{n}}
\end{array}\right.,
\label{cv}
\end{eqnarray}
and
\begin{eqnarray}
c_A=\left \{\begin{array}
{r@{\quad\quad}l}
-\frac{1}{2} & {\nu},{{p}}\\
\frac{1}{2} & e,{{n}}
\end{array}\right..
\label{ca}
\end{eqnarray}
Here $W^{\mu \nu}(q)$ and $S_\ell(p)$ stand for the $W$-boson
propagator in a magnetic field and charged lepton propagator in
presence of a magnetized plasma. The forms of the propagators are
given in \cite{Bhattacharya:2002aj, D'Olivo:2002sp, Erdas:1998uu,
Bravo Garcia:2007uc,Bhattacharya:2004nj}. The $Z^{\mu \nu}(q)$ is the
$Z$-boson propagator in vacuum and $S_{\nu_\ell}(p)$ is the neutrino
propagator in a thermal bath of neutrinos. The coefficients $a$, $b$,
$c$ are given as:
\begin{eqnarray}
a &=& a_W + a_T + a_Z\,,
\label{at}\\
b &=& b_W + b_T + b_Z\,,
\label{bt}\\
c &=& c_W + c_T + c_Z\,.
\label{ct}
\end{eqnarray}
In Eq.~(\ref{at}) $a_W$, $a_T$ and $a_Z$ are composed of the parallel
and perpendicular parts as shown in Eq.~(\ref{sigmanb}).  For the case
of a charge symmetric plasma, which perhaps existed in the early
universe we have:
\begin{eqnarray}
b &=&  -\frac{4 g^2 k_0}{3 M_W^2 M_Z^2}  
\langle E_{\nu^{\rm B}_\ell}\rangle N_{\nu_\ell}\nonumber\\ 
&-& \frac{2 e {\mathcal B}
g^2}{M_W^4} 
\int_0^\infty \frac{dp_3}{(2\pi)^2}
\sum_{n=0}^\infty \sum_{\lambda=\pm 1}
\left[\frac{k_3}{E_{\ell,\,n}}\left( p_3^2  +
\frac{m_\ell^2}{2}\right)\delta^{n,0}_{\lambda,1} +
k_0 E_{\ell,\,n}\right]f_{\ell,\,n}\,,
\label{bcompls}\\
c &=& - \frac{2e {\mathcal B} g^2}{M_W^4} \int_0^\infty \frac{dp_3}{(2\pi)^2}
\sum_{n=0}^\infty \sum_{\lambda=\pm 1}
\left[k_0\left(E_{\ell,\,n} - \frac{m_\ell^2}{E_{\ell,\,n}}\right)
\delta^{n,0}_{\lambda,1} +
\frac{k_3 p_3^2}{E_{\ell,\,n}}\right]f_{\ell,\,n} \,.
\label{ccompls}\\
a_\perp &=& -\frac{2 g^2 e {\mathcal B}}{M_W^4} 
\int_0^\infty \frac{dp_3}{(2\pi)^2}
\sum_{n=0}^\infty \sum_{\lambda=\pm 1} \left(\frac{\mathcal H}{2E_{\ell,\,n}}
+ \frac{m_\ell^2}{E_{\ell,\,n}}\right) f_{\ell,\,n}
+\frac{g^2}{3M_W^4}\langle E_{\nu^{\rm B}_\ell}\rangle N_{\nu_\ell}\,,
\label{aperpfs}\\
a_\parallel &=&  -\frac{2 g^2 e {\mathcal B}}{M_W^4} 
\int_0^\infty \frac{dp_3}{(2\pi)^2}
\sum_{n=0}^\infty \sum_{\lambda=\pm 1} 
\frac{m_\ell^2}{E_{\ell,\,n}}f_{\ell,\,n}
+\frac{g^2}{3M_W^4}\langle E_{\nu^{\rm B}_\ell}\rangle N_{\nu_\ell}\,.
\label{aparafs}
\end{eqnarray}
In the above equations,
\begin{eqnarray}
{\mathcal H}= e {\mathcal B}(2 n +1 -\lambda)\,. 
\label{hf}
\end{eqnarray}
where $n$ is a positive integer including zero and $\lambda$ can take
only two values $\pm 1$. The $n$ corresponds to the Landau level
number occurring in the energy of the charged leptons in a magnetic
field and $\lambda$ corresponds to the spin states of the leptons. The
charged lepton wave-functions and the dispersion relation of them are
calculated in \cite{Bhattacharya:2004nj, Bhattacharya:2002qf, 
Bhattacharya:2007vz}. More over,
\begin{eqnarray}
f_{\ell,\,n} = \frac{1}{e^{\beta(E_{\ell,\,n} - \mu_\ell)}+1}\,,\,\,&&
\bar{f}_{\ell,\,n} = \frac{1}{e^{\beta(E_{\ell,\,n} + \mu_\ell)}+1}\,,
\label{pdistrb}\\
N_\ell = \frac{e {\mathcal B}}{2\pi^2}\sum_{n=0}^\infty \sum_{\lambda=\pm 1}
\int_0^\infty dp_3 f_{\ell,\,n}\,,
\,\,&&
\bar{N}_\ell = \frac{e {\mathcal B}}{2\pi^2}\sum_{n=0}^\infty 
\sum_{\lambda=\pm 1}\int_0^\infty dp_3 
\bar{f}_{\ell,\,n}\,,
\label{nnbar}
\end{eqnarray}
and $N^0_\ell$ and $\bar{N}^0_\ell$ corresponds to $N_\ell$ and
$\bar{N}_\ell$ with $E_{\ell,\,n}$ in the distribution functions
replaced by $E_{\ell,\,0}$, that is $N^0_\ell$ and $\bar{N}^0_\ell$
are the particle and anti-particle number densities in the lowest
Landau level. The symbol $\delta^{n,0}_{\lambda,1}=1$ only when $n=0$
and $\lambda=1$ and zero in other cases. $N_{\nu_\ell}$ and ${\bar
N}_{\nu_\ell}$ are the number densities of the neutrinos and
antineutrinos. Here $E_{\nu^{\rm B}_\ell}=p\cdot u$ is the energy of
the background neutrinos and $\langle E_{\nu^{\rm B}_\ell}\rangle$ is
the average energy per unit volume per neutrino in the heat bath From
the expressions of $a$, $b$, $c$ we immediately notice that all the
contributions in the charge symmetric case are proportional to
$M_W^{-4}$ or $G_F^2$. In addition to the charged leptons in the
medium we can also have neutrons and protons in it, in that case the
calculations will get extra contributions from diagrams where the
protons replace the charged leptons and the neutrons replace the
neutrino in the the self-energy diagrams.  In the ${\mathcal B} \to 0$
limit the last term on the right hand side of the above equation
vanishes and we have,
\begin{eqnarray}
b_{{\mathcal B}\to 0}=- \frac{16\sqrt{2} G_F k_0}{3 M_Z^2}  
\langle E_{\nu^{\rm B}_\ell}\rangle N_{\nu_\ell} - 
\frac{16 \sqrt{2} G_F k_0}{M_W^2} 
\langle E_{\ell} \rangle \,N_\ell\,.
\label{nr}
\end{eqnarray}
The Eq.~(\ref{nr}) resembles the results found in \cite{notr}.  As the
coefficients $c$ and $a_\parallel-a_\perp$ are directly related to the
existence of a non-zero magnetic field their zero field correspondence
is not strictly permitted due to the non-perturbative nature of the
Landau quantization of the charged fermions in presence of a magnetic
field.

To order of $g^2$ the dispersion relation becomes,
\begin{eqnarray}
E_{\nu_\ell} = \left[|{\bf k}|^2 - 2 c {\bf k}\cdot {\hat{\bf b}} 
+ 2(a_\parallel - a_\perp) k_\perp^2 \right]^{1/2} +b\,,
\end{eqnarray}
where we have taken the positive sign of the square root in
Eq.~(\ref{mndisp}).  The above dispersion relation can be simplified
by binomially expanding the square root and neglecting terms of order
more than $g^2$. The expansion gives,
\begin{eqnarray}
E_{\nu_\ell} &=& |{\bf k}| -  c {\hat{\bf k}}\cdot {\hat{\bf b}}
+ (a_\parallel - a_\perp) \frac{k_\perp^2}{|{\bf k}|} +b\,,\nonumber\\
&=& |{\bf k}| -  c \cos \theta 
+ (a_\parallel - a_\perp) {|{\bf k}|} \sin^2 \theta +b\,,
\label{fdispr}
\end{eqnarray}
where $k^3 = k_z = |{\bf k}| \cos \theta$. The above equation implies
that in presence of a magnetized medium the effective-potential acting
on the neutrinos is of the form,
\begin{eqnarray}
V_{\rm eff} = b -  c \cos \theta
+ (a_\parallel - a_\perp) {|{\bf k}|} \sin^2 \theta \,.
\label{veff}
\end{eqnarray}
From the expressions of $a_\parallel$ and $a_\perp$ in the {\bf CP}
symmetric case we see that in the lowest Landau level $a_\parallel -
a_\perp$ is zero and in that case the effective potential is
independent of $a$. With the form of the effective potential in
Eq.~(\ref{veff}) the problem of neutrino oscillations in the {\bf CP}
symmetric magnetized plasma in the early universe can be tackled.
From Eq.~(\ref{ccompls}) we see that the first term on the right hand
side of the equation vanishes trivially in the zero field limit. 
\section{Conclusion}
\label{conclu}
The talk was focussed on the calculation of the self-energy of the
neutrino in a medium seeded with a uniform classical magnetic
field. The calculations were carried out in the unitary gauge where
the unphysical Higgs contribution does not appear. The background is
supposed to be comprised of the charged leptons and neutrinos
equillibriated at the same temperature. The magnitude of the magnetic
field is such that only linear contributions of the field appear in
the charged $W^{\pm}$ boson propagators but all orders of the field
are present in the charged lepton propagators. In the present
calculation we have assumed that the neutrinos are also in a state of
thermal equilibrium.  We explicitly write down the possible form of
the neutrino self-energy in a magnetized medium by applying the
concepts of Lorentz symmetry. The form of the self-energy becomes
involved when the Lorentz breaking contributions are taken into
account.  The calculation is specially aimed at trying to find out the
order $G_F^2$ contributions to the neutrino self-energy. This specific
order of the coupling is important as we see that in a charge
symmetric plasma the neutrino self-energy is proportional to $G_F^2$
only.
\vskip .5cm
\noindent{\bf Acknowledgement}: The research work discussed in the
talk is partially supported by DGAPA-UNAM project IN119405 and
Conacyt, Mexico, grant No. 52975.

\end{document}